\documentclass[aps,amsmath,amssymb,reprint,superscriptaddress]{revtex4-2}

\usepackage{graphicx}
\usepackage{dcolumn}% Align table columns on decimal point
\usepackage{bm}% bold math

\usepackage{subfigure}

\usepackage[colorlinks,linkcolor=blue,anchorcolor=blue,citecolor=blue]{hyperref}

\begin{document}
\title{YbNi$_4$Mg: Superheavy fermion with enhanced Wilson ratio and magnetocaloric effect}

\author{Xiaoci Zhang}
\affiliation{Beijing National Laboratory for Condensed Matter Physics, Institute of Physics, Chinese Academy of Sciences, Beijing 100190, China}
\affiliation{School of Physical Sciences, University of Chinese Academy of Sciences, Beijing 100049, China}

\author{Te Zhang}
\affiliation{Beijing National Laboratory for Condensed Matter Physics, Institute of Physics, Chinese Academy of Sciences, Beijing 100190, China}
\affiliation{School of Physical Sciences, University of Chinese Academy of Sciences, Beijing 100049, China}

\author{Zhaotong Zhuang}
\affiliation{Beijing National Laboratory for Condensed Matter Physics, Institute of Physics, Chinese Academy of Sciences, Beijing 100190, China}
\affiliation{School of Physical Sciences, University of Chinese Academy of Sciences, Beijing 100049, China}

\author{Zixuan Leng}
\affiliation{Beijing National Laboratory for Condensed Matter Physics, Institute of Physics, Chinese Academy of Sciences, Beijing 100190, China}
\affiliation{School of Physical Sciences, University of Chinese Academy of Sciences, Beijing 100049, China}

\author{Zixuan Wei}
\affiliation{Beijing National Laboratory for Condensed Matter Physics, Institute of Physics, Chinese Academy of Sciences, Beijing 100190, China}
\affiliation{School of Physical Sciences, University of Chinese Academy of Sciences, Beijing 100049, China}

\author{Xinyang Liu}
\affiliation{Beijing National Laboratory for Condensed Matter Physics, Institute of Physics, Chinese Academy of Sciences, Beijing 100190, China}

\author{Junsen Xiang}
\affiliation{Beijing National Laboratory for Condensed Matter Physics, Institute of Physics, Chinese Academy of Sciences, Beijing 100190, China}
\affiliation{School of Physical Sciences, University of Chinese Academy of Sciences, Beijing 100049, China}

\author{Shuai Zhang}
\affiliation{Beijing National Laboratory for Condensed Matter Physics, Institute of Physics, Chinese Academy of Sciences, Beijing 100190, China}
\affiliation{School of Physical Sciences, University of Chinese Academy of Sciences, Beijing 100049, China}

\author{Peijie Sun}
\email{pjsun@iphy.ac.cn}
\affiliation{Beijing National Laboratory for Condensed Matter Physics, Institute of Physics, Chinese Academy of Sciences, Beijing 100190, China}
\affiliation{School of Physical Sciences, University of Chinese Academy of Sciences, Beijing 100049, China}
\affiliation{Songshan Lake Materials Laboratory, Dongguan, Guangdong 523808, China}

\date{\today}

\begin{abstract}
A comprehensive study of the low-temperature properties of YbNi$_4$Mg has revealed evidence of a superheavy-fermion state, characterized by a large electronic specific-heat coefficient $\gamma_0$\,$\approx$\,5.65 J\,mol$^{-1}$\,K$^{-2}$ and an elevated Wilson ratio $R_W$\,=\,32.1. No magnetic ordering was observed down to 70\,mK; however, a broad maximum appears in the specific heat at $T^*$\,=\,0.3 K, along with a shoulder in the derivative of susceptibility d$\chi$/d$T$ and resistivity d$\rho$/d$T$. These features indicate a cooperative yet short-ranged magnetism entwined with the superheavy Fermi liquid. The large Wilson ratio, which is also detected in other superheavy-fermion compounds lacking long-range order, might be a signature of residual spin fluctuations. Applying a weak magnetic field of $\sim$0.1 T induces a metamagnetic-like crossover, as demonstrated by the quasi-adiabatic demagnetization measurements showing a broad minimum in the temperature-field trace. Here, an enhanced magnetocaloric cooling effect stemming from the field-sensitive superheavy-fermion state is observed, rivaling that of the well-established insulating magnetic coolants like the rare-earth garnet Gd$_3$Ga$_5$O$_{12}$.  

\end{abstract}

%\keywords{}
\maketitle

\section{INTRODUCTION}

Strongly correlated electron systems on the verge of long-range magnetic order offer a rich avenue for exploring emergent quantum phases. Central to these systems are heavy-fermion compounds characterized by enhanced electronic specific-heat coefficient $\gamma_0$ (=\,$C/T$($T$$\rightarrow$0)) \cite{si10}. While typical heavy-fermion compounds exhibit $\gamma_0$ ranging from $10^2$ to $10^3$ mJ\,mol$^{-1}$\,K$^{-2}$ \cite{ste84}, a small number of Kondo-lattice materials display exceptionally high values of $\gamma_0$\,$\sim$\,10$^4$ mJ\,mol$^{-1}$K$^{-2}$. These materials are referred to as superheavy-fermion (SHF) compounds \cite{sereni18,shimura20,mun13,yats96,sereni18phyb}, which often lack magnetic order but manifest a broad specific-heat maximum or shoulder at several hundred miliKevin. 

SHF compounds typically exhibit a weak Kondo effect because, according to the single-channel Kondo impurity model \cite{rajan83}, the Kondo temperature $T_K$ (=\,1.29$\pi$$R$/6$\gamma_0$, with $R$ being the gas constant) is inversely proportional to $\gamma_0$. Likewise, the Ruderman–Kittel–Kasuya–Yosida (RKKY) exchange interaction is similarly weak, because it depends on the exchange coupling $J_{cf}$ between local $f$-electrons and conduction bands, much like the Kondo effect. In a Doniach-type magnetic phase diagram \cite{doniach77}, these weak-coupling Kondo systems are expected to magnetically order at low temperature, with the RKKY interaction dominating over the Kondo effect. However, most SHF compounds studied thus far reveal no long-range magnetic order. Structurally, these compounds often possess a face-centered cubic (fcc) lattice, where edge-sharing tetrahedrons form a three-dimensional frustrated structure. This suggests that typical SHF phenomena might also rely on a specific structural arrangement to ensure a large spin degeneracy at the ground state.  

Beyond their fundamental research significance, SHF materials can offer practical benefits for adiabatic demagnetization refrigeration (ADR) in the sub-Kelvin regime, potentially serving as alternatives \cite{jang15,tokiwa16,sereni20} to classical hydrated paramagnetic salts \cite{giau73}. In SHF materials, the huge electronic specific heat at very low temperatures, coupled with its field tunability, provides a novel way to harness significant magnetic entropy by field tuning the many-body Kondo entanglement \cite{tokiwa16}. In a fundamental perspective, the large magnetocaloric effect (MCE) in SHF compounds arises from highly-degenerate electronic states, unlike the low-energy magnetic excitations in paramagnetic salts. Additionally, SHF materials, being metals, may exhibit high thermal conductivity, offering a notable advantage over the traditional paramagnetic salts that are more thermally insulating \cite{amb55}. This advancement is of significant practical interest given the global shortage of $^4$He and $^3$He cryogenic gases \cite{cho2009}. 

YbNi$_4$Mg belongs to a large materials family with fcc structure (${\rm MgCu_{4}Sn}$-type, space group F$\overline{4}$3m). The sole previous work on this compound \cite{Linsinger2011}, which measured a poly crystal sample, shows a Curie-Weiss (CW) behavior in a wide temperature range of 20$-$300 K with the effective moment of free Yb$^{3+}$. The magnetic susceptibility increases drastically upon cooling down to 2 K, with no signature of long-range magnetic order. These results indicate weak Kondo coupling and RKKY interactions in YbNi$_4$Mg, placing it close to a magnetic instability. Notably, some of its Yb-based homologues crystallizing in fcc lattice, like YbCu$_4$Ni \cite{sereni18}, are already known to exhibit SHF phenomenon. 

\begin{table*}[]
\caption{\label{tab:table1}
The superheavy-fermion properties of some representative materials discussed in this work.  The space group (S.G.) and the nearest neighbour distance of rare-earth ions, $d_{nn}$, are also included. $T^*$ marks the temperature of the broad maximum in specific heat, typically characterizing a short-rang order (or long-range order, if indicated separately). $T_K$, $A$, $\gamma_0$ and $\chi_0$ are the Kondo temperature, the coefficient of the $T$-square resistivity, the Sommerfeld values of specific-heat coefficient and susceptibility at zero temperature limit, respectively. $R_W$ represents the calculated Wilson ratio and $R_{KW}$ the Kadowaki-Woods ratio. $A$, $\gamma_0$, $\chi_0$ and $R_{KW}$ are represented in unit $\mu\Omega$\,cm\,K$^{-2}$, J\,mol$^{-1}$K$^{-2}$, emu\,mol$^{-1}$ and $\mu\Omega$\,cm\,(K\,mol/mJ)$^2$, respectively. These compounds typically crystallize in fcc structure, adopting either the full Heusler Cu$_2$MnAl-type ($Fm\overline{3}m$), the half Heusler-type ($F\overline{4}3m$), the MgCu$_4$Sn-type with space group $F\overline{4}3m$, or their closely related derivatives. Additionally, a canonical heavy fermion CeCu$_6$ is also included in this table for comparison. } 
\begin{ruledtabular}
\begin{tabular}{lcccccccccr} 
Materials & S.G.& $d_{nn}$ (\,\AA)&  $T^*$ (K) & $T_K$ (K) & $\gamma_0$ & $\chi_0$   & $R_W$ & $A$ & $R_{KW}$ & ref. \\
\colrule
YbNi$_4$Mg & $F\overline{4}3m$ &4.9817 & 0.3 & 0.9 & 5.65 & 2.49 & 32.1  & 22.17 & 0.7$\times$10$^{-6}$ & this work
\\
YbCu$_4$Ni & $F\overline{4}3m$ &4.95 & 0.21 & 0.84 & 7.5 & 1.2 & 11.7 & 11.43 & 0.2$\times$10$^{-6}$ &\cite{sereni18,osato24,shimura2022magnetic}
\\
YbPt$_2$Sn & $P6_{3}/mmc$\footnote{The hexagonal ZrPt$_2$Al-type structure of YbPt$_2$Sn is a close variant of the fcc structure adopted by YbPt$_2$In.} &4.49 & 0.25 & $-$ & 10 &  5.68\footnote{Magnetic susceptibility data at 0.4 K is used.} & 41.5 & $-$& $-$&\cite{gruner14,jang15}
\\
YbPt$_2$In & $Fm\overline{3}m$&4.71 & 0.18 & $-$ & 15 & $-$ & $-$ & $-$& $-$ &\cite{gruner14}
\\
YbPd$_2$In & $Fm\overline{3}m$ & 4.71& 0.21 &$-$& 10 & $-$& $-$ & $-$& $-$  &\cite{Gas19}
\\
Ce$_4$Pt$_{12}$Sn$_{25}$ & $Im\overline{3}$ &6.14 & 0.19 ($T_N$) & 1.2 & 10\footnote{$\gamma_0$\,=\,10 J\,mol$^{-1}$K$^{-2}$ is read from the $C/T$ value at 90 mK, in accordance to the temperature where $\chi_0$ is read.} & 0.31 & 2.3 & 11 & 5.0$\times$10$^{-5}$\footnote{For calculating $R_{KW}$, $\gamma_0$ = 0.45 J\,mol$^{-1}$K$^{-2}$ at the 0\,K limit is employed \cite{lee10}. } &\cite{kurita10,lee10}
\\
YbCo$_2$Zn$_{20}$ & $Fd\overline{3}m $ & 6.06 & $-$ &1.5 & 7.9 & 0.53\footnote{$\chi_0$ is obtained from the measurement made at 60 mK, 0.3 T \cite{honda14}.}  & 4.9 & 165 & 2.7$\times$10$^{-6}$ &\cite{honda14,tori07,shimura11,takeuchi11, take11,shimura20} 
\\ 
YbBiPt & $F\overline{4}3m$ & 4.66 & 0.4 ($T_N$) & 1.0 &8.0 & $-$ & 2.64 & $-$ & $-$&\cite{fisk91,mun13}
\\
CeNi$_9$Ge$_4$ & $I4/mcm$\footnote{The tetragonal structure is a variant of cubic NaZn$_{13}$ structure.} & 5.64 & $-$& 3.0 & 5.5& 0.13 & 1.73 & $-$ & $-$ &\cite{killer04}
\\
PrInAg$_2$ & $Fm\overline{3}m$ & 5.00 & $-$ & 0.8 & 7.0 & 0.04 & 0.42 & $-$ & $-$ &\cite{yats96,movs99}
\\
CeCu$_6$ & Pnma &4.85 & $-$ & 3.9 & 1.6 & 0.03\footnote{$\chi_0$ of CeCu$_6$ is taken as the average of the susceptibilities measured along the three principle axes at 100 mK.} & 1.37 & 120 & 4.7$\times$10$^{-5}$ &\cite{schr92,amato87}
\\
\end{tabular}
\end{ruledtabular}
\end{table*}

In this work, we have successfully synthesized high-quality single crystals of YbNi$_4$Mg for the first time. Our low-temperature study of the magnetic, transport and thermodynamic properties down to below 0.1 K unveils a SHF state characterized by an enormous value of $\gamma_0$\,=\,5.65 J\,mol$^{-1}$\,K$^{-2}$. This state is entwined with a cooperative paramagnetism below $T^*$\,$\approx$\,0.3 K, which features a large Wilson ratio $R_W$\,=\,32.1. Applying magnetic field reveals a metamagnetic-like crossover near 0.1 T, where an enhanced  MCE is observed, surpassing that of typical paramagnetic salts like Gd$_3$Ga$_5$O$_{12}$ (GGG). Our results are compared to those of other intermetallic compounds ever recognized as SHF materials, with some key material properties summarized in Tab.~\ref{tab:table1}. Evidently, these SHF materials generally crystallize in fcc lattice or closely related derivatives and are characterized by notably weak Kondo coupling. For typical SHF materials lacking long-range magnetic order, a large Wilson ratio seems common, which might offer an important insight into the novel magnetic ground state, potentially involving metallic spin liquid.  

\section{EXPERIMENTAL DETAILS}

Single crystals of YbNi$_{4}$Mg was grown by Mg self-flux method. High-purity Yb and Ni were first mixed in a 1:4 molar ratio and pressed into a metal block under argon atmosphere. The metal block was then loaded into an alumina crucible, sealed in a quartz tube under vacuum, and heated to 1100\,$^{\circ}$C for a dwell time of two days. The resulting YbNi$_4$ ingot and Mg shot (in a molar ratio 1:11) were sealed into a tantalum tube in argon atmosphere and heated to 1075\,$^{\circ}$C. After keeping this temperature for 12 hours, the mixture was slowly cooled down to 750\,$^{\circ}$C over 12 days. Following a rapid cooling to room temperature, the crucible with sample melt was taken out, resealed in quartz tube, and centrifuged at 750$^{\circ}$C to separate the single crystals (Fig.~\ref{fig:XRD} inset) from the Mg flux. The nonmagnetic reference of poly crystal LuNi$_4$Mg was prepared by sintering the constituent elements at 1500\,$^{\circ}$C in a tantalum tube.   

X-ray diffraction was performed for both the finely ground powder and the as-grown facet of YbNi$_{4}$Mg single crystals to characterize the crystal structure. Electrical resistivity was measured using the standard four-probe method, while specific heat was assessed by the thermal relaxation method down to below 0.1 K in a $^3$He-$^4$He dilution refrigerator installed in a physical property measurement system (PPMS, Quantum Design). DC magnetic susceptibility and magnetization measurements were carried out using a vibrating sample magnetometer equipped with a SQUID sensor (SQUID-VSM), initially down to 1.8 K and further extended to 0.4 K by integrating  a ${\rm ^{3}He}$ insert (iHelium3). 

The MCE was evaluated in a homemade quasi-adiabatic thermal stage adapted for an Oxford $^3$He cryostat. The thermal stage features a 3$\times$3 mm sapphire plate suspended by thin nylon wires in a G10 glassfiber frame (to be sketched in Fig.~\ref{MCE-20240604}(a) inset). A thermometer (CX-1010 bare chip sensor, Lake Shore Cryotronics Inc.) fixed to the sapphire plate, connected with manganin wires ($\phi$\,=\,25 $\mu$m), was employed to monitor the temperature variation during the MCE measurements. Due to the excellent thermal insulation between the thermal stage and the cold bath, the lowest accessible initial temperature $T_i$ of the loaded sample remains 0.8 K when a 27 mg YbNi$_4$Mg sample is used, much higher than the base temperature (0.28 K) of the $^3$He cryostat. To further evaluate the practical cooling effect, we have also constructed a cooling device based on the commercial PPMS sample puck (to be shown Fig.~\ref{MCE-20240604}(c) inset), which includes a GGG shielding layer and the coolant material (YbNi$_{4}$Mg) supported by Vespel straws, as described previously \cite{xiang2024giant}. This device enabled us to measure the quasi-adiabatic cooling effect of a larger amount of coolant starting from the $T_i$\,=\,2 K environment of PPMS.   

\section{Experimental results}

\begin{figure}[]
\centering
\includegraphics[width=0.96\linewidth]{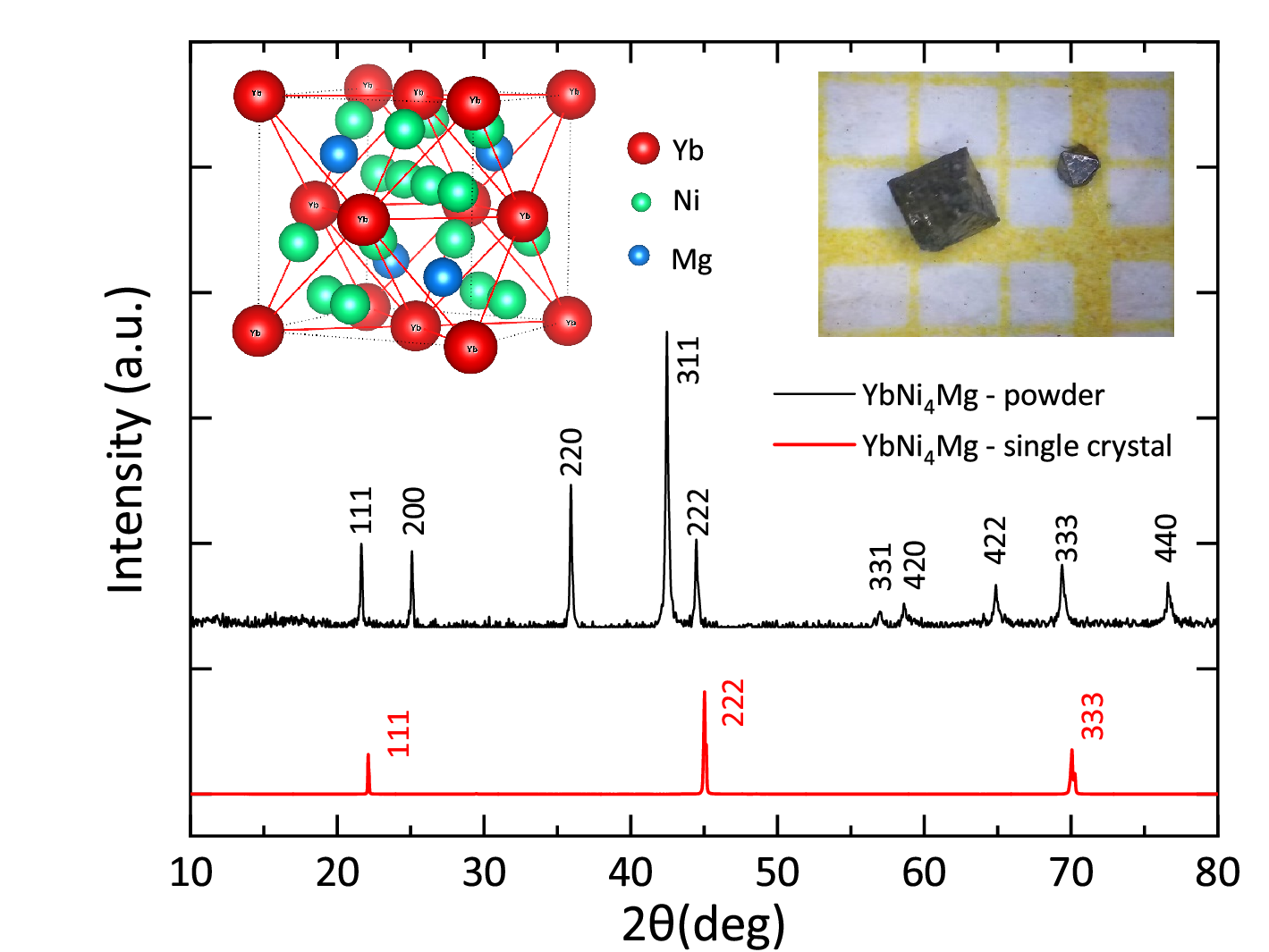}
\caption{\label{fig:XRD} X-ray diffraction patterns of the finely ground powder and an as-grown single crystal facet. All measured Bragg peaks are indexed according to the MgCu$_4$Sn-type structure. Insets: the fcc lattice (left) and the optical image of the YbNi$_4$Mg single crystals (right). }  
\end{figure}

The powder x-ray diffraction results, shown in Fig.~\ref{fig:XRD}, confirm the ${\rm MgCu_{4}Sn}$-type structure with all Bragg peaks properly indexed. The obtained lattice parameter, $a$\,=\,7.0452\,\AA\,, is in reasonable agreement with the literature value of 7.0107\,\AA\, \cite{Linsinger2011}. Furthermore, the x-ray pattern measured from the single crystal facet shows selected Bragg peaks consistent with the (111) crystallographic plane. As illustrated in Fig.~\ref{fig:XRD} inset, the fcc lattice features a network of edge-sharing tetrahedra formed by ${\rm Yb^{3+}}$ ions, creating a three-dimensional analogue of the frustrated triangular lattice. Interestingly, most of the prototypical SHF compounds crystallize in the fcc lattice, as summarized in Tab.~\ref{tab:table1}. Consistent with other SHF materials, the nearest Yb-Yb distance within the tetrahedron for YbNi$_4$Mg measures 4.9817\,\AA, significantly larger than that in typical Yb-based heavy fermion compounds like YbRh$_2$Si$_2$ ($\sim$4.0\,\AA). 

\begin{figure*}
\includegraphics[width=0.98\linewidth]{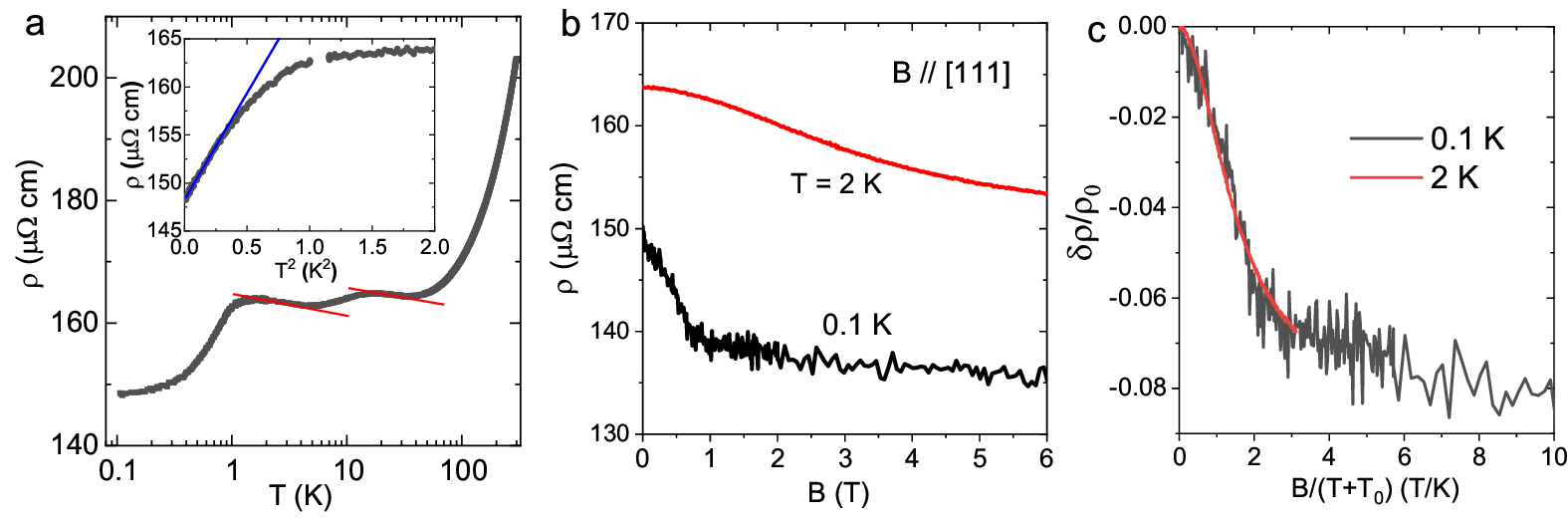}
\caption{\label{fig:RT01} (a) Electrical resistivity $\rho(T)$ of ${\rm YbNi_{4}Mg}$ measured down to 0.1 K, with current applied within the (111) plane. Red lines indicate the $-$ln$T$ dependence. Inset shows $\rho$ vs $T^2$, where the low-$T$ part exhibits a linear variation (see the blue line) with a slope $A$\,=\,22.17 $\mu\Omega$\,cm\,K$^{-2}$. (b) Isothermal magnetoresistivity $\rho(B)$ measured at $T$\,=\,0.1 and 2 K. (c) $\delta$$\rho$/$\rho_0$ ($\delta$$\rho$\,=\,$\rho$$-$$\rho_0$) as a function of the scaled field, $B$/($T$+$T_0$). The two curves collapse into one by assuming a characteristic temperature $T_0$\,=\,0.25 K.}
\end{figure*}

The $T$-dependent resistivity $\rho(T)$ of YbNi$_4$Mg, shown in Fig.~\ref{fig:RT01}(a), is characteristic of Kondo-lattice compound by showing two distinct maxima. The high-$T$ one at $\sim$\,20 K results from the crystal electric field (CEF) splitting of the Yb$^{3+}$ $J$\,=\,7/2 Hund's rule multiplet, while the low-$T$ one at $\sim$1.5 K marks the onset of Kondo-singlet coherence. The $-$ln$T$-dependent $\rho(T)$, which reflects local Kondo scattering, is clearly observed above the two maxima (red lines in Fig.~\ref{fig:RT01}(a)). The low-$T$ resistivity, depicted as $\rho$ vs $T^2$ (Fig.~\ref{fig:RT01}(a) inset), displays a linear dependence below about 0.5 K, consistent with the Fermi-liquid description of $\rho$\,=\,$\rho_0$\,+\,$A$\,$T^2$, with $A$\,=\,22.17 $\mu\Omega$\,cm\,K$^{-2}$.

Fig.~\ref{fig:RT01}(b) shows the isothermal magnetoresistivity $\rho(B)$ measured at $T$ = 0.1 K and 2 K. At 0.1 K, $\rho(B)$ exhibits a strong initial decrease with increasing field up to about 1 T, changing into a more gradual decline at $B$\,$>$\,1 T. This contrasts with $\rho$($B$) measured at $T$\,=\,2 K, which decreases smoothly in the whole field range. However, as shown in Fig.~\ref{fig:RT01}(c), $\delta$$\rho$/$\rho_0$ ($\delta$$\rho$\,=\,$\rho$$-$$\rho_0$) for the two distinctly different temperatures collapse into a single curve when $B$ is scaled by temperature, $B$/($T$+$T_0$), consistent with theoretical predictions for the Kondo impurity model \cite{schl83}. Here, $T_0$ is generally a positive number characterizing the strength of antiferromagnetic Kondo coupling: $T_0$\,=\,0.25 K obtained for YbNi$_4$Mg verifies the significantly weak Kondo effect in this compound. Note that, if ferromagnetic correlations dominate, $T_0$ becomes negative \cite{petri00}. Absence of a $\delta$$\rho(B)$ peak, typically associated with the suppression of a magnetically ordered phase as seen in YbRh$_2$Si$_2$ \cite{frie10}, suggests that YbNi$_4$Mg lacks long-range magnetic order at least down to 100 mK.

\begin{figure} [tb]
	\centering
	\includegraphics[width=0.97\linewidth]{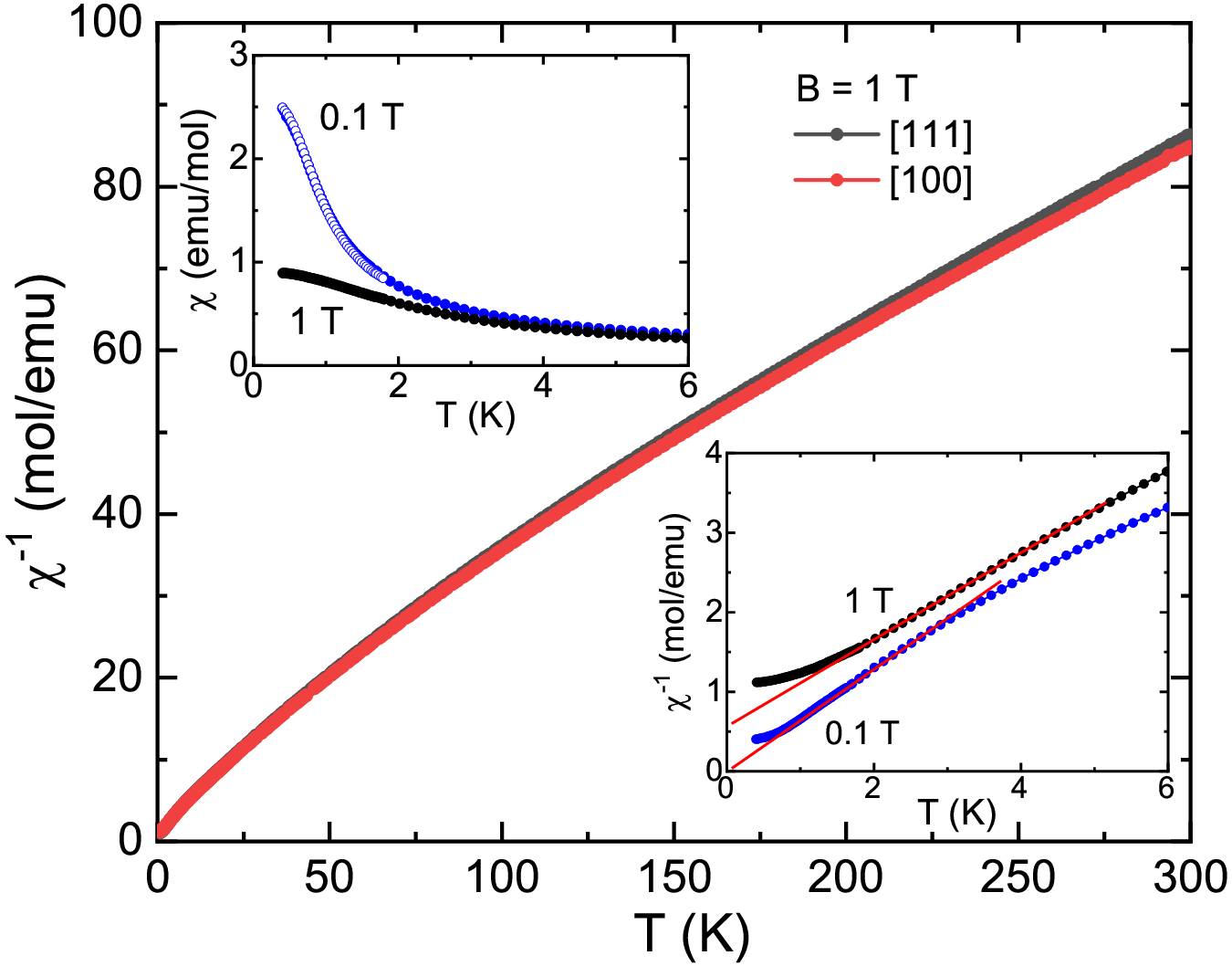}
	\caption{Inverse susceptibility $\chi^{-1}(T)$ measured in field $B$\,=\,1 T along [100] and [111] axes. The two curves are nearly identical, suggestive of a weak magnetic anisotropy. Upper inset: Low-$T$ $\chi(T)$ measured down to 0.4 K with $B$\,=\,0.1 and 1 T applied along [111]. The $\chi(T)$ measured in field-cooling (open circle) and zero-field-cooling (closed circle) for $B$\,=\,0.1 T are practically identical. Lower inset: low-$T$ $\chi^{-1}(T)$ with linear CW fittings delineated by red lines. 
	}
	\label{fig:M} 
\end{figure}

\begin{figure}[tb]
\includegraphics[width=0.98\linewidth]{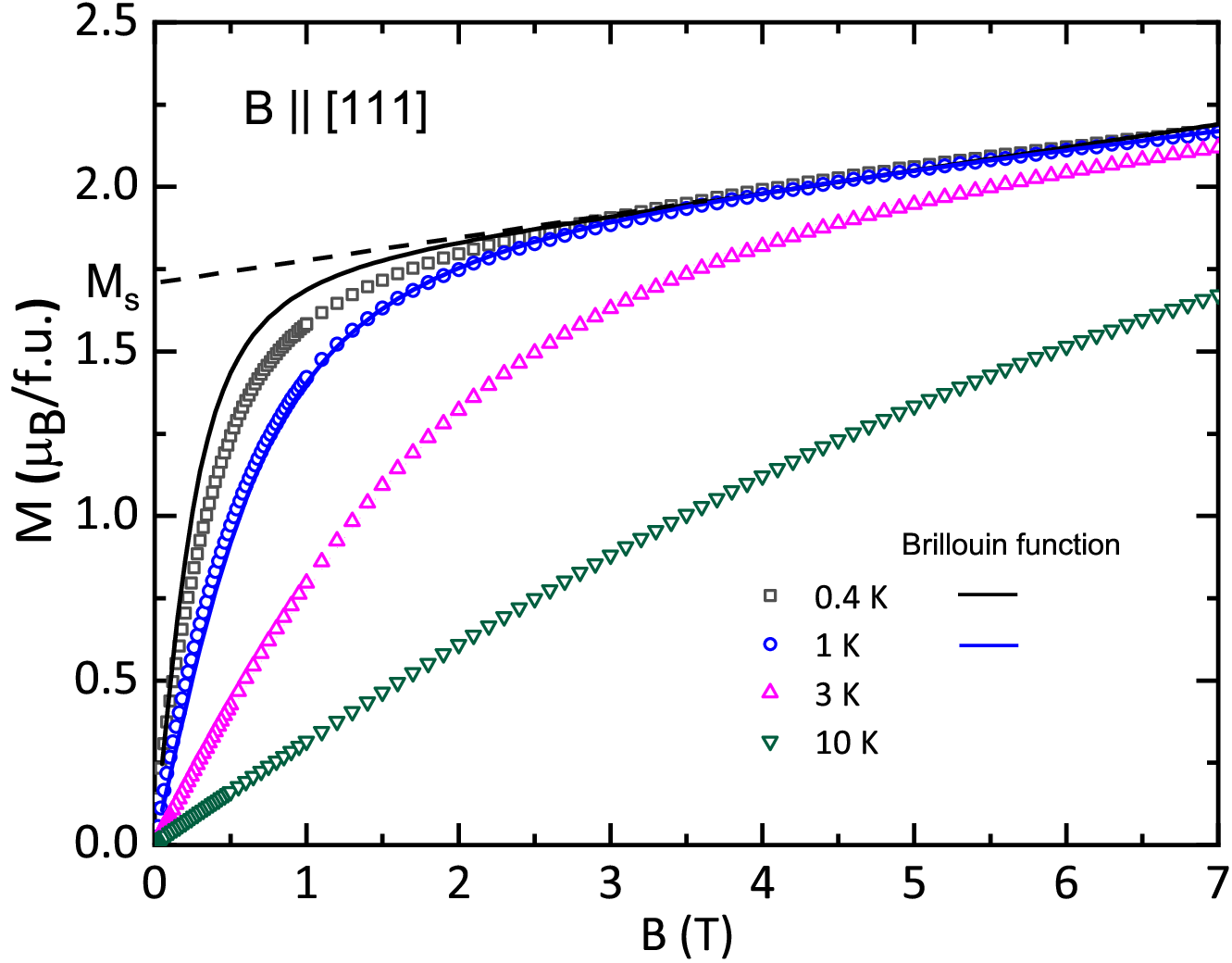}
\caption{\label{fig:fitting of MT} Isothermal magnetization $M(B)$ measured at various temperatures for $B$\,$\parallel$\,[111]. Linearly extrapolating the $M(B)$ data measured at 0.4 K for $B$\,$>$\,2 T to zero field yields a saturate moment $M_s$\,$\approx$\,1.7 $\mu_B$. Solid lines represent the calculations made for a non-interacting spin system based on the Brillouin function, including a field linear term with slopes of 0.07 $\mu_{B}$/(T f.u.) (0.4 K) and 0.06 $\mu_{B}$/(T f.u.) (1.0 K) to account for the van Vleck contribution.}
\end{figure}

The inverse susceptibility $\chi^{-1}(T)$ measured in an external field $B$\,=\,1 T along [100] and [111] axes (Fig.~\ref{fig:M}, main panel) exhibits a weak curvature, with negligible anisotropy. A modified CW law, $\chi$\,=\,$C_0$/($T$\,$-$\,$\theta_p$)\,+\,$\chi_0$, was employed to analyze the data in the range 100$-$300 K. Here, $C_0$ =\,($N\mu_{eff}^2$/3$k_B$), $\theta_p$, and $\chi_0$ represent the Curie constant, the paramagnetic Weiss temperature, and a $T$-independent susceptibility, respectively. Our analysis yielded an effective moment $\mu_{eff}^{HT}$\,=\,4.51 $\mu_{B}$, close to that of free Yb$^{3+}$ ion, 4.54 $\mu_{B}$. The superscript HT indicates that these parameters are from the high-$T$ (HT) analysis, distinguishing them from those obtained in the subsequent low-$T$ (LT) analysis. The high-$T$ Weiss temperature $\theta_p^{HT}$\,=\,$-$6.43 K is primarily influenced by thermally populated CEF states rather than the exchange interaction, which is anticipated to be weaker given the absence of magnetic order. $\chi_0$ is also an important fitting parameter and it was set to $\chi_0$\,=\,0.00385 emu/mol in this case. This additional paramagnetic contribution is likely due to an enhanced van Vleck susceptibility resulting from the weak CEF splitting in YbNi$_4$Mg, as discussed in relation to the Schottky anomaly observed in specific heat. 

The low-$T$ (0.4$-$6 K) susceptibility $\chi(T)$ and its inverse $\chi(T)^{-1}$ are shown in Fig.~\ref{fig:M} upper and lower insets, respectively. Results for both $B$\,=\,0.1 T and 1 T are included to examine the field dependence of the exchange interaction, which reveal no signature of long-range magnetic ordering. For $B$\,=\,0.1 T, $\chi(T)$ was measured under both field-cooling and zero-field-cooling conditions, and no difference is observed between them. This indicates the absence of spin freezing down to at least 0.4 K. A CW analysis of the low-$T$ $\chi(T)^{-1}$ shows that the Weiss temperature $\theta_p^{LT}$ is small and field dependent. For $B$\,=\,1 T, $\theta_p^{LT}$ is $-$0.9 K, while for $B$\,=\,0.1 T, $\theta_p^{LT}$ becomes practically zero within the measurement uncertainties and the CW fitting yields $\mu_{eff}^{LT}$\,=\,3.15 $\mu_B$ (Fig.~\ref{fig:M} lower inset), indicative of weak and field-sensitive exchange interaction. 

The weak exchange interaction is also evidenced by the magnetization isotherm $M(B)$ displayed in Fig.~\ref{fig:fitting of MT}. At the lowest accessible temperature for magnetic measurements, $T$\,=\,0.4 K, $M(B)$ increases rapidly at low fields ($B$\,$<$\,1 T), before transitioning to a weak and quasi-linear increase at higher fields. Linearly extrapolating the $M(B)$ line at $B$\,$>$\,2 T to zero field yields a saturate moment $M_s$\,$\approx$\,1.7 $\mu_B$ ($B$$\parallel$[111]), which is primarily dictated by the ground-state doublet. This value is much smaller than the theoretical saturate magnetization of a free Yb$^{3+}$ ion, $g_J$$J$\,=\,4\,$\mu_{B}$, but close to $g_J^{LT}$$J_{eff}$\,=\,1.82 $\mu_B$, with the effective Land\'{e} g-factor $g_J^{LT}$\,=\,3.64 (obtained from the value of $\mu_{eff}^{LT}$\,=\,$g_J^{LT}$$\sqrt{J_{eff}(J_{eff} + 1)}$\,$\mu_B$) and the effective angular moment $J_{eff}$\,=\,1/2 of the ground state doublet. In line with the weak antiferromagnetic exchange interactions, the measured $M(B)$ curve for $T$\,=\,0.4 K shows slightly reduced values near 1 T compared to the calculation based on the Brillouin function for a non-interacting spin system. Upon warming up to 1 K, such discrepancy disappears, with the Brillouin function reasonably reproducing the measured $M(B)$.  The fits also require a field-linear term $\alpha$$B$, with $\alpha$\,=\,0.07 $\mu_{B}$/(T f.u.) (0.4 K) and 0.06 $\mu_{B}$/(T f.u.) (1.0 K), to account for the van Vleck paramagnetism arising from the weak CEF splitting.

\begin{figure}
\includegraphics[width=0.95\linewidth]{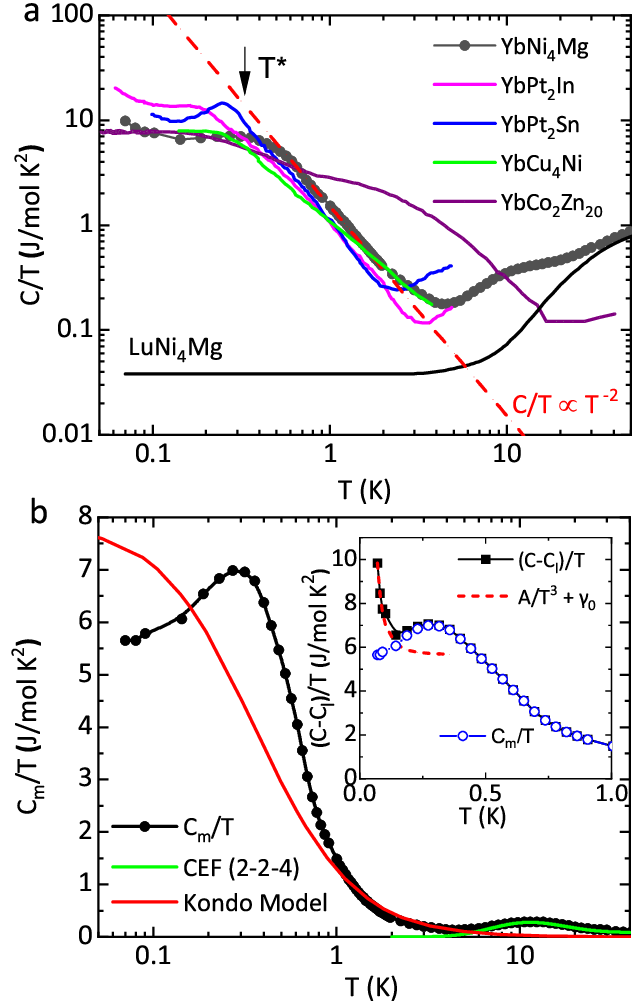}
\caption{\label{HC-nuclear} (a) $C(T)/T$ of YbNi$_4$Mg and its nonmagnetic analog LuNi$_4$Mg, depicted in a double-logarithmic representation, is compared with that of several other SHF compounds (YbCu$_4$Ni \cite{sereni18}, YbCo$_2$Zn$_{20}$ \cite{shimura20,take11}, YbPt$_{2}$Sn and YbPt$_{2}$In \cite{gruner14}). The dashed line indicates the $C/T$$\sim$$T^{-2}$ dependence for $T$\,$>$\,$T^*$. (b) The magnetic specific heat  $C_{m}/T$ is compared with $C_{\rm K}$/$T$ calculated for a spin-1/2 Kondo model (see text). The green solid line is a numerical fit to the Schottky anomaly observed around 10 K, based on a CEF scheme of doublet-doublet-quartet (2-2-4). Inset: The low-$T$ upturn of ($C-C_l$)/$T$ is fitted to $A_n$/$T^3$\,+\,$\gamma_0$, i.e., the sum of a nuclear contribution and a constant term of heavy fermions, with $A_n$ = 1.438\,$\times$\,10$^{-3}$ J\,K\,mol$^{-1}$ and $\gamma_0$\,=\,5.65 J\,mol$^{-1}$\,K$^{-2}$. 
}
\end{figure}

\begin{figure}[tb]
\includegraphics[width=0.98\linewidth]{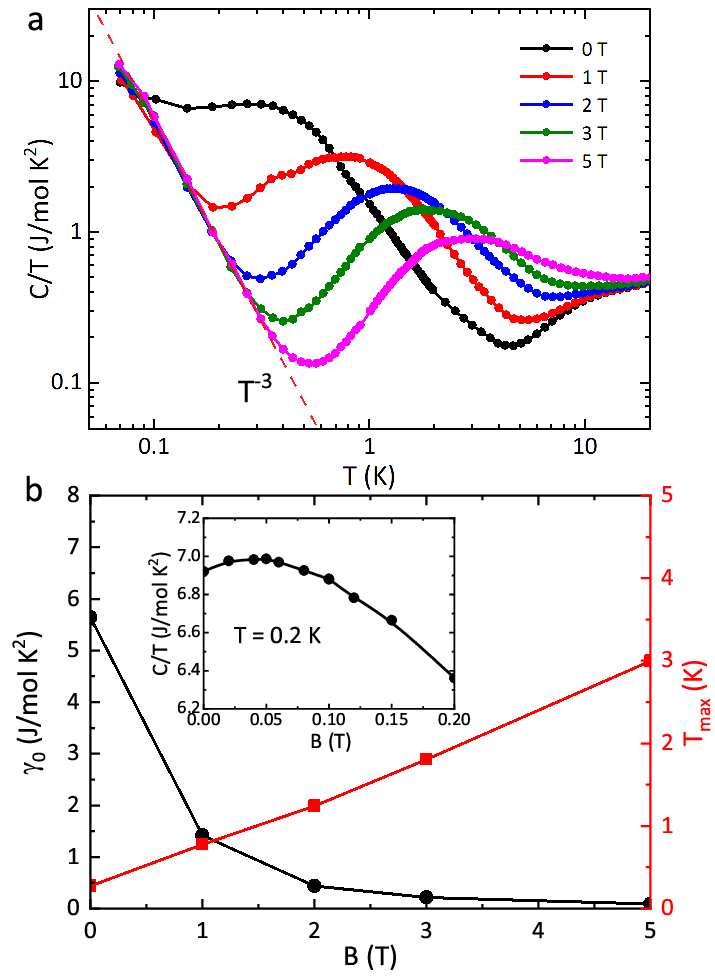}
\caption{\label{fig:CoverT} (a) $C(T)/T$ measured in selected fields ($B$\,$\parallel$\,[111]). The $T^{-3}$ dependence of the low-$T$ $C/T$ upturn (the dashed red line) indicates the high-$T$ tail of the nuclear Schottky contribution, which gradually extends to higher temperatures with applying field. (b)  The Sommerfeld coefficient $\gamma_0(B)$, and $T_{max}(B)$, which marks the temperature of the broad $C/T$ maximum related to Zeeman splitting. Inset: $C(B)/T$ measured at 0.2 K in a narrow field interval.}
\end{figure}

\begin{figure}[tb]
\includegraphics[width=0.98\linewidth]{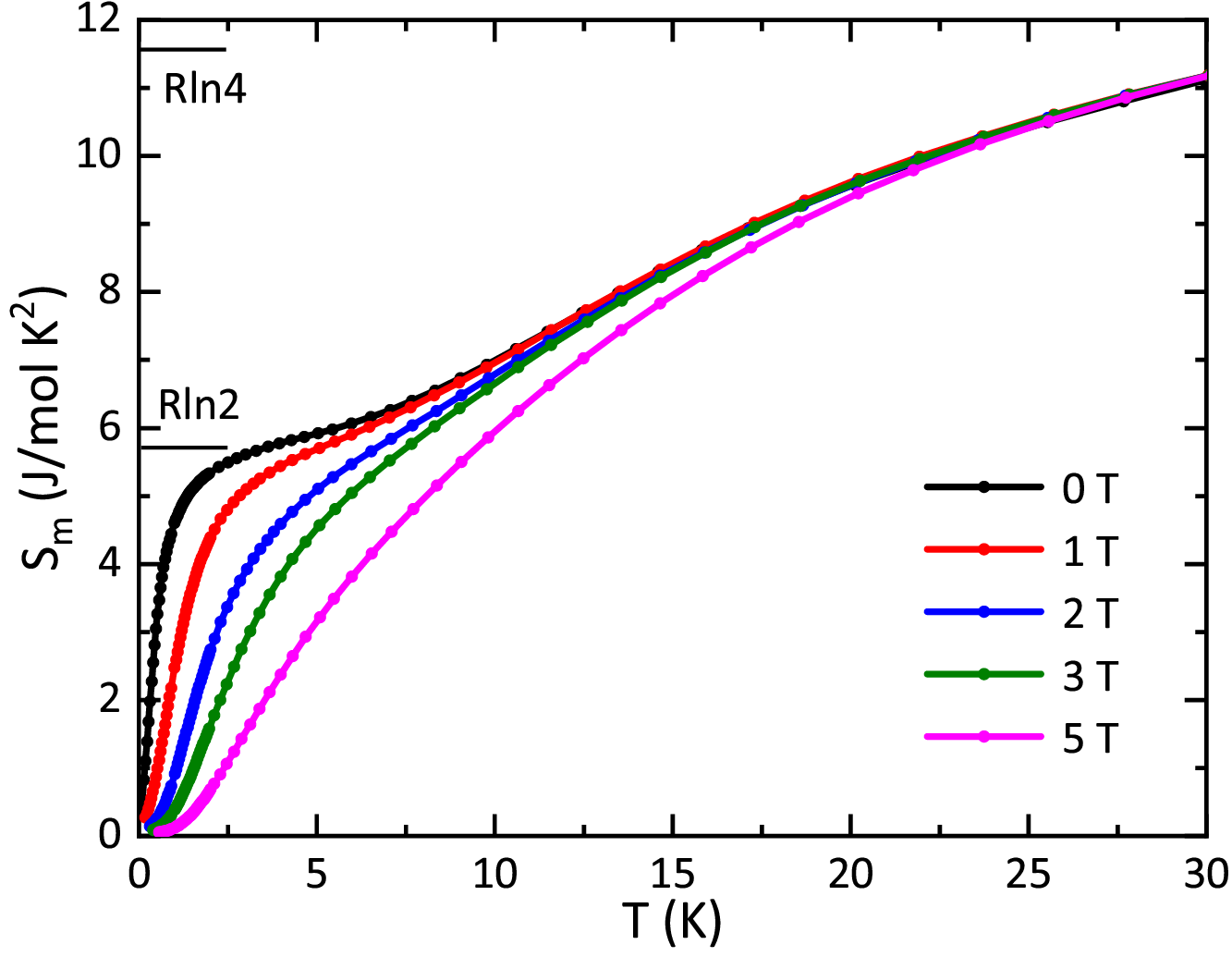}
\caption{\label{fig:Magnetic-entropy} The magnetic entropy $S_m(T)$ estimated by integrating $C_m/T$ with respect to temperature. }
\end{figure}

The specific heat divided by temperature, $C/T$, is plotted in Fig.~\ref{HC-nuclear}(a) for both YbNi$_4$Mg and its nonmagnetic analogue LuNi$_4$Mg. In YbNi$_4$Mg, $C/T$ is drastically enhanced upon cooling below about 3 K following a power-law dependence $C/T$\,$\sim$\,$T^{-2}$ (see the dashed red line). This behavior persists down to $T^*$\,$\approx$\,0.3 K, where a broad maximum indicative of the onset of short-range order is observed. This $C/T$ profile shares similarity with those of several other Yb-based SHF compounds, including YbCu$_4$Ni \cite{sereni18}, YbPt$_{2}$Sn and YbPt$_{2}$In \cite{gruner14}, which are also shown in Fig.~\ref{HC-nuclear}(a). The power-law dependence at $T$\,$>$\,$T^*$ is common in SHF systems and has been ascribed to short-range spin correlations that are too weak to induce long-range magnetic order \cite{jang15,gruner14}. Furthermore, a large $C/T$ upturn below about 0.1 K is observed, due to the Schottky anomaly of Yb nuclear magnetic moment exposed in the hyperfine magnetic field produced by 4$f$ electrons \cite{loun62}. This upturn, therefore, strongly corroborates the formation of short-range order in this compound. As a comparison, such a nuclear Schottkey anomaly is absent in YbCo$_2$Zn$_{20}$ down to below 0.1 K (see Fig.~\ref{HC-nuclear}(a)), and correspondingly, no broad $C/T$ maximum signaling short-range order is observed. Instead of the low-$T$ $C/T$\,$\sim$\,$T^{-2}$ contribution as discussed above, YbCo$_2$Zn$_{20}$ exhibits an enhanced $C/T$ that spans over a much wider temperature range up to above 10\,K, due to its nearly degenerate CEF ground and excited doublets, which are separated by only 6$-$9 K \cite{takeuchi11, take11,shimura20}.

To approximate the magnetic specific heat $C_m$, we first subtract $C$ of LuNi$_{4}$Mg, which serves as the lattice contribution $C_l$. Next, we fit the low-$T$ upturn of ($C$$-$$C_l$)/$T$ below 0.1 K, as shown in Fig.~\ref{HC-nuclear}(b) inset, to $\gamma_0$\,+\,$A_n$\,$T^{-3}$, representing the sum of a heavy-electron contribution and a nuclear Schottky term. This analysis yields $\gamma_0$\,=\,5.65 J\,mol$^{-1}$\,K$^{-2}$ and $A_n$\,=\,1.438 $\times$ 10$^{-3}$ J\,K\,mol$^{-1}$. Fig.~\ref{HC-nuclear}(b) main panel shows $C_m$/$T$, obtained as ($C$$-$$C_l$)/$T$$-$$A_n$\,$T^{-3}$, with the lattice and nuclear parts removed from the measured specific heat. Markedly, $C_m$/$T$ exhibits a broad maximum at $T^*$. This behavior contrasts with the theoretical $C_{\rm K}/T$ curve calculated for a spin-1/2 Kondo model \cite{DESGRANGES1982} with $T_K$\,=\,0.9 K (to be determined below), which increases smoothly upon cooling and saturates at a much larger Sommerfeld coefficient as $T$$\rightarrow$0. Moreover, the $C_{\rm K}/T$ enhancement upon cooling is too weak to explain the $C/T$\,$\sim$\,$T^{-2}$ dependence observed at $T$\,$>$\,$T^*$.  

Fig.~\ref{fig:CoverT}(a) displays the low-$T$ $C(T)/T$ measured in selected fields ($B$\,$\parallel$\,[111]). A dashed line ($C/T$\,$\propto$\,$T^{-3}$) is drawn to highlight the nuclear specific heat observed in all fields. With applying field, the broad maximum at 0.3 K ($B$\,=\,0) due to the onset of short-range order transitions to an even broadened maximum resulting from the Zeeman-split quasi-particle band. 
The value of $\gamma_0$, obtained via the fitting procedure described above, decreases drastically with field, whereas $T_{max}$, which marks the position of $C/T$ maximum, increases almost linearly (Fig.~\ref{fig:CoverT}(b) main panel). As shown in Fig.~\ref{fig:CoverT}(b) inset, a careful measurement of $C(B)/T$ at 0.2 K in a narrow field interval (0$-$0.2 T) reveals a broad maximum at $\sim$0.05 T, evidencing the suppression of the short-range magnetism and SHF state.

Fig.~\ref{fig:Magnetic-entropy} shows the magnetic entropy $S_m$ estimated by integrating $C_m/T$ with respect to temperature. It levels off near 4 K for $B$\,=\,0, reaching the full entropy ($R$ln2) of the lowest-lying Kramers doublet. Based on the $S_m(T)$ results and according to Desgranges and Schotte‘s calculation for spin-1/2 Kondo model \cite{desg82}, $S_{m}(T_K)$\,=\,0.65\,$R$ln2, we obtain $T_K$\,$\approx$\,0.9 K for YbNi$_4$Mg. Upon further warming, the $S_m(T)$ traces tend to approach $R$ln4 at about 30 K, involving the magnetic entropy of two low-lying doublets. Correspondingly, a broad $C_m$/$T$ maximum around 10 K is recognized (see Fig.~\ref{HC-nuclear}(b)) as a Schottcky anomaly arising from thermal population of the low-lying CEF 4$f$ states. For a Yb$^{3+}$ ion in cubic site symmetry, the $J$\,=\,7/2 multiplet splits into two doublets and one quartet. As shown in Fig.~\ref{HC-nuclear}(b), a three-level CEF scheme (2$-$2$-$4) with $\Delta_{1}$\,=\,40 K and $\Delta_{2}$\,=\,150 K can reasonably reproduce the broad maximum at $\sim$10 K, consistent with the entropy results in Fig.~\ref{fig:Magnetic-entropy}. 

\begin{figure}[tp]
\includegraphics[width=0.96\linewidth]{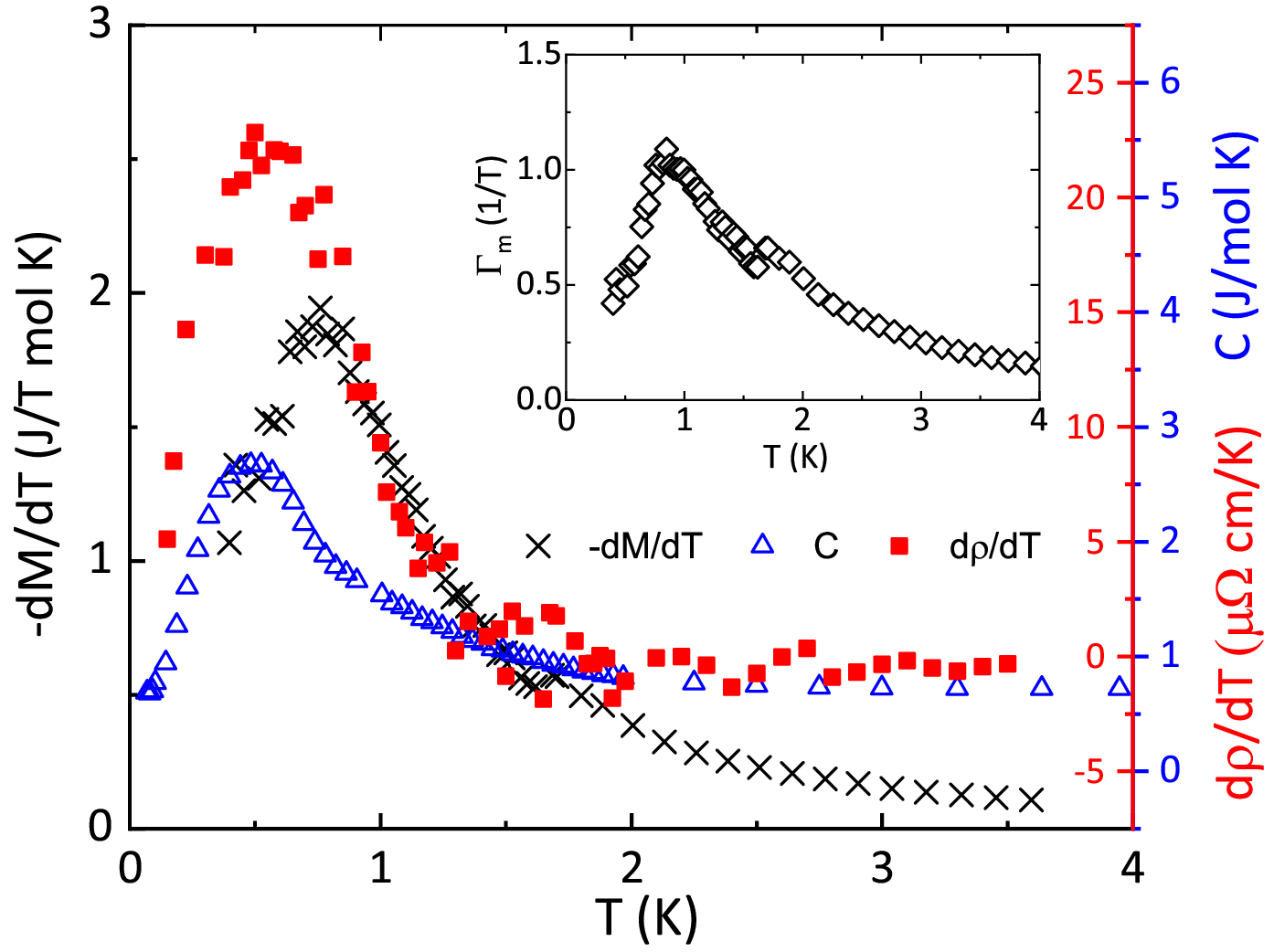}
\caption{\label{fig:grun-20240604} Comparison between the negative magnetization derivative $-$d$M$/d$T$ (measured in $B$\,=\,0.1 T), the specific heat $C$, and the resistivity derivative d$\rho$/d$T$. Inset shows the magnetic Gr\"uneisen ratio $\Gamma_m$ calculated from the ratio of the former two quantities.
}
\end{figure}

\begin{figure*}
\includegraphics[width=0.99\linewidth]{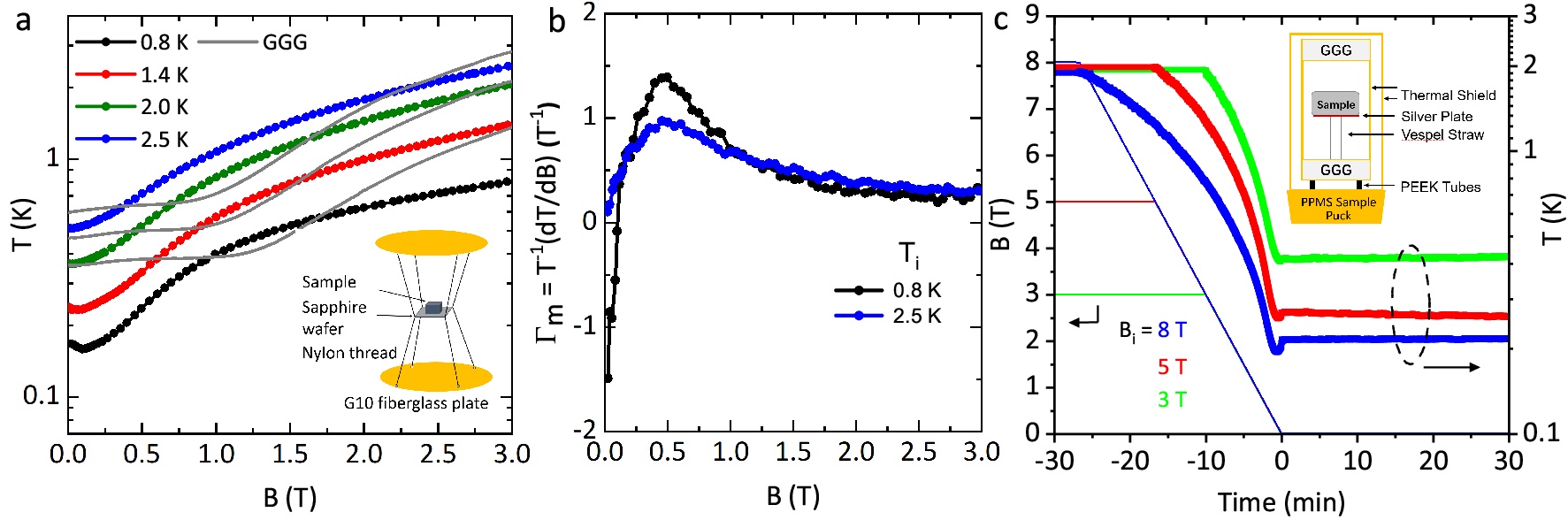}
\caption{\label{MCE-20240604} (a) Quasi-adiabatic demagnetization $T(B)$ traces measured by employing a home-made thermal stage loaded in an Oxford $^3$He refrigerator. The measurements were performed from a fixed initial field $B_i$\,=\,3 T and different $T_i$'s, with the field ramping rate 0.3 T\,min$^{-1}$. A total of about 27 mg single crystal samples were used. Inset: The thermal stage used for recording $T(B)$ consists of a sapphire wafer suspended by nylon thin thread ($\phi$\,=\,0.1 mm) on a pair of G10 plates. (b) Gr\"uneisen ratio $\Gamma_M(B)$ calculated from selected $T(B)$ traces with $T_i$ = 0.8 and 2.5 K.  (c) Demagnetization cooling shown as temperature and field versus time, achieved in a quasi-adiabatic thermal device equipped in the commercial PPMS (see inset to this panel and ref.\cite{xiang2024giant}). In this measurement, a collection of YbNi$_4$Mg single crystals (total mass 160 mg) was placed on a silver stage, which is supported by Vespel straws and surrounded by GGG and an additional gold-plated copper shield to reduce the thermal leak from environment. }
\end{figure*}

Fig.~\ref{fig:grun-20240604} provides a further examination of the SHF state by comparing the low-$T$ $C$, $-$d$M$/d$T$, and d$\rho$/d$T$. According to the Maxwell relation, d$M$/d$T$ (=\,d$S$/d$B$) detects the field derivative of entropy, and is related to $C$, which measures $T$(d$S$/d$T$). As shown in Fig.~\ref{fig:grun-20240604}, both $-$d$M$/d$T$ and $C$, as well as their ratio $\Gamma_m$\,=\,$-$($dM$/$dT$)/$C$ that defines the magnetic Gr\"uneisen ratio (Fig.~\ref{fig:grun-20240604} inset) \cite{tokiwa09, zhu03}, strongly increase upon cooling, assuming a maximum below 1 K. Significantly, the $\Gamma_m(T)$ profile deviates from the low-$T$ saturation behavior that is expected for typical nonmagnetic heavy-fermion state, suggestive of enhanced spin fluctuations near $T^*$ associated with the short-range ordering. Fig.~\ref{fig:grun-20240604} also displays d$\rho$/d$T$, which serves as a transport indicator of magnetic entropy through spin-fluctuation-enhanced scatterings \cite{fisher68}. The maxima observed in all these quantities versus $T$ $-$ though broad and occurring at slightly different temperatures $-$ provide strong evidence of magnetic entropy accumulation below 1 K, albeit the absence of magnetic ordering. These behaviors are characteristic of a weak-coupling Kondo system with weak exchange interactions, and possibly indicate a certain kind of cooperative magnetism with residual fluctuations.  

Next, we study the MCE by recording $T(B)$ in a quasi-adiabatic demagnetization process, which provides direct insight into the magnetic entropy landscape associated with spin fluctuations. As delineated in Experimental Details, we utilize a home-designed quasi-adiabatic thermal stage within a $^3$He cryostat to study this effect (see Fig.~\ref{MCE-20240604}(a) inset). Fig.~\ref{MCE-20240604}(a) main panel displays the obtained $T(B)$ traces of a 27 mg sample measured from an initial field $B_i$\,=\,3 T and various $T_i$'s, revealing a strong magnetic cooling effect. Starting from $T_i$\,=\,2 K, a final temperature $T_f$\,=\,0.36 K is reached by the end of the demagnetization. Markedly, the $T(B)$ trace for $T_i$\,=\,0.8 K exhibits a broad minimum at $B^*$\,$\approx$\,0.1 T, indicating enhanced quantum fluctuations near this characteristic field. Here, a pronounced temperature minimum $T_{min}$\,=\,0.16 K is obtained. This $T(B)$ minimum is in reasonable agreement with the broad $C(B)/T$ maximum shown in Fig.~\ref{fig:CoverT}(b) inset. The substantial MCE roots in the small energy scales of $T_K$ and $T^*$, which allow for a full release of the magnetic entropy associated with the ground-state doublet ($R$\,ln2) below a couple of Kelvins (Fig.~\ref{fig:Magnetic-entropy}), while being easily tunable by small fields. Similar $T(B)$ behavior with a broad temperature minimum at low field has been also observed in other SHF compounds like YbPt$_2$Sn ($B^*$\,$\approx$\,0.1 T) \cite{jang15} and YbCo$_2$Zn$_{20}$ ($B^*$\,$\approx$\,0.6 T) \cite{tokiwa16}. No magnetic quantum critical point is expected at $B^*$ due to the absence of long-range magnetic order in these compounds.  

For comparison, Fig.~\ref{MCE-20240604}(a) also shows the $T(B)$ results measured for a single crystal of GGG under similar initial conditions. In contrast to YbNi$_4$Mg, the final temperature $T_f$'s for GGG settle down much higher temperatures after demagnetization (for instance, $T_f$\,=\,0.46 K for $T_i$\,=\,2 K). While $T(B)$ for YbNi$_4$Mg continuously decreases from $B_i$ down to $B^*$\,=\,0.1 T, the $T(B)$ traces for GGG flatten out below a considerably higher field of $\sim$1 T, reflecting its much stronger short-range ordering that sets in at a higher temperature (0.8 K) \cite{fisher73}.  

From the experimentally obtained $T(B)$ results, the magnetic Gr\"uneisen ratio as a function of field, $\Gamma_m(B)$\,=\,$-(dM/dT)/C$\,=\,$T^{-1}(\partial{T}/\partial{B})$, can be readily calculated. $\Gamma_m(B)$ is a direct measure of MCE and it generally changes sign when $B$ is tuned across a quantum critical point, where $\Gamma_m(B)$ diverges \cite{zhu03,garst05}. As shown in Fig.~\ref{MCE-20240604}(b), the calculated $\Gamma_m(B)$ for $T_i$\,=\,2.5 K and 0.8 K reveals a broad maximum at $B$\,$\sim$\,0.5 T. At $B$\,$<$\,$0.5$ T, $\Gamma_m(B)$ for the lower $T_i$ of 0.8 K smoothly decreases, revealing a sign change at 0.1 T. No signature or precursor of $\Gamma_m(B)$ divergence can be observed near the sign change. These behaviors are characteristic of SHF state, which is inherently field-sensitive, and similar to those of YbCo$_2$Zn$_{20}$ \cite{tokiwa16}, where a smooth sign change of $\Gamma_m(B)$ (without divergence) is observed at its metamagnetic crossover field $B^*$\,$\approx$\,0.6 T. Note that, YbCo$_2$Zn$_{20}$ is a naive SHF material lacking short-range order \cite{take11}. Likewise, many other nonmagnetic heavy-fermion compounds also feature weak metamagnetic crossover, as observed in CeCu$_6$ ($B^*$\,=\,1.7 T) \cite{schr92,aoki99}. Markedly, it has been noted that the crossover field $B^*$ in heavy-fermion materials commonly scales to the Kondo temperature $T_K$ \cite{honda14}.  
 
To verify the cooling ability of YbNi$_4$Mg as a potential ADR coolant, we have also fabricated a quasi-adiabatic demagnetization cooling device based on the standard sample puck of commercial PPMS, as described in Experimental Details and sketched in Fig.~\ref{MCE-20240604}(c) inset. Here, GGG crystals were used to serve as a thermal shielding layer because when starting from similar initial conditions, it can cool down to a higher but still similar temperature compared to YbNi$_4$Mg (see Fig.~\ref{MCE-20240604}(a)). For this device we use a relatively large amount of YbNi$_4$Mg sample (160 mg) and precool it to $T_i$\,=\,2 K via helium gas in the PPMS sample chamber. Starting from $B_i$\,=\,8 T, $T_f$ reaches as low as 0.21 K (lower than $T^*$) for a holding time of more than 1 hour, comparable to the cooling performance of the SHF compound YbCu$_4$Ni \cite{shimura2022magnetic}. For $B_i$\,=\,3 T, $T_f$\,=\,0.42 K is obtained. Reflecting the metamagnetic crossover at $B^*$, a shallow $T(B)$ valley is also visible before $T_f$ is reached. These results indicate a strong MCE effect of SHF that deserves further attention for practical applications in sub-Kelvin refrigeration. 

\section{discussion and conclusion}

As shown in Tab.~\ref{tab:table1}, SHF compounds mostly have large spacing between rare-earth ions, fcc structure, low short-range ordering temperature (if any) and low Kondo temperature that are typically close to or less than 1 K. To differentiate their behavior from canonical heavy-fermion characteristics as observed in CeCu$_6$ (also included in Tab.~\ref{tab:table1}), these compounds are further examined by comparing their susceptibility and specific heat. From the magnetic susceptibility $\chi_0$ and the magnetic specific-heat coefficient $\gamma_0$\,=\,$C_m$/$T$ at zero-temperature limit, one can estimate the Wilson ratio, $R_{\rm W}$\,=\,($\pi^2$$k_B^2$/3\,$\mu_B^2$)\,$\chi_0$/$\gamma_0$. This ratio serves as a convenient measure of spin correlations in Fermi liquid and is employed here as an additional probe of the ground state property. For a free electron gas, $\chi_0$\,=\,$\mu_B^2$\,$g_F$ and $\gamma_0$\,=\,($\pi^2$/3)$k_B^2$\,$g_F$, resulting in $R_{\rm W}$\,=\,1. Here, $g_F$ is the electronic density of states at the Fermi level. For strongly correlated heavy-fermion systems, however, $R_{\rm W}$\,=\,2 is theoretically expected \cite{hewson93} and typically observed in experiments \cite{fisk87}. 

For YbNi$_4$Mg, $\gamma_0$ is 5.65 J\,mol$^{-1}$\,K$^{-2}$ and $\chi_0$ = 2.49 emu\,mol$^{-1}$ (measured at 0.4 K and in 0.1 T, see Fig.~\ref{fig:M} upper inset). This results in $R_{W}$ = 32.1, signalling the more strongly enhanced $\chi_0$ compared to $\gamma_0$. Although a high Wilson ratio in strongly correlated electron systems is often associated with ferromagnetic correlations, such as in YbRh$_2$(Si$_{0.95}$Ge$_{0.05}$)$_2$ with $R_W$\,=\,30 (0.065 T) \cite{gegen05}, our results clearly demonstrate antiferromagnetic interactions in YbNi$_4$Mg, as indicated by the negative value of $\theta_p^{\rm LT}$ and the positive value of $T_0$ obtained in describing $\delta \rho(B)$. Intriguingly, a notably large value of $R_{W}$ appears rather common among the SHF compounds so-far studied (Tab.~\ref{tab:table1}). For example, in YbCu$_4$Ni, the susceptibility and specific heat measured down to below 100 mK \cite{sereni18} yield $\gamma_0$\,=\,7.5 J mol$^{-1}$\,K$^{-2}$ and $\chi_0$\,=\,1.2 emu\,mol$^{-1}$, leading to $R_{W}$\,$\approx$\,11.7. For YbPt$_2$Sn, a large Wilson ratio $R_{W}$\,=\,41.5 can be deduced based on the published experimental results \cite{gruner14,jang15}. Likewise, $R_W$\,$\sim$\,4.9 is obtained for YbCo$_2$Zn$_{20}$ (Tab.~\ref{tab:table1}), which is moderately enhanced compared to the Fermi liquid value. 

However, not all SHF compounds reveal large $R_W$.  YbBiPt has $R_W$\,=\,2.64 \cite{fisk91} and Ce$_4$Pt$_{12}$Sn$_{25}$ has $R_W$\,=\,2.3 (Tab.~\ref{tab:table1}), in reasonable agreement with that expected for Fermi liquid. These two compounds are distinct from other SHF compounds by showing long-range antiferromagnetic order at $T_N$\,=\,0.4 K and 0.19 K, respectively, suggesting that a large Wilson ratio is probably a feature only in SHF compounds lacking long-range magnetic order. Note, however, that PrInAg$_2$ is exceptional: its large $\gamma_0$ arises from quadrupolar rather than spin fluctuations, resulting in a non-enhanced $R_W$\,=\,0.42 (Tab.~\ref{tab:table1}).   

To help understand the large Wilson ratios observed in typical SHF compounds without attributing them to ferromagnetic interactions, we note that a large $R_W$ of 35 has been experimentally observed in a quantum spin liquid candidate Na$_4$In$_3$O$_8$, too. The large $R_W$ has been theoretically explained by the enhanced spin fluctuations in its quantum spin liquid phase with strong spin-orbital coupling \cite{chen13}. Likewise, SHF compounds generally feature frustrated spins and weak Kondo coupling, with uncompensated moments persisting down to very low temperatures. Although being metals, they may have an exotic ground state akin to spin liquid, accounting for the frequently observed short-range magnetic order entwined with the SHF state. Whether the enhanced $R_W$ in SHF compounds indicates spin-liquid-like fluctuations remains an interesting question for future study. 

With the enhanced Wilson ratios in mind, we have also estimated the Kadowaki-Woods (KW) ratio $R_{\rm KW}$ (=\,$A$/$\gamma_0^2$) of YbNi$_4$Mg and related compounds to examine the relationship between transport and thermodynamics of the SHF state. Based on the $A$ coefficient of the $T$-square resistivity and the value of $\gamma_0$, we calculate $R_{\rm KW}$\,=\,0.7$\times$10$^{-6}$ $\mu\Omega$\,cm\,(K\,mol/mJ)$^2$ for YbNi$_4$Mg (see Tab.~\ref{tab:table1}). 
This value is one order of magnitude smaller than the typical KW ratio for Yb- or Ce-based heavy-fermion compounds with CEF doublet ground state, i.e., $R_{\rm KW}$\,=\,1$\times$10$^{-5}$ $\mu$$\Omega$\,cm(K\,mol/mJ)$^2$ \cite{tsujii05}. A similarly reduced $R_{\rm KW}$ is also observed for YbCu$_4$Ni and YbCo$_2$Zn$_{20}$ (see Tab.~\ref{tab:table1}). Most probably, this reduction is related to the residual spin fluctuations in SHF state, which strongly enhances $\gamma_0$ beyond Kondo picture, while being less influential to resistivity. Such a reduced KW ratio has been observed in low-dimensional Kondo lattices with enhanced spin fluctuations, too \cite{lyu21}.  

In conclusion, the distinct signatures of Kondo effect and the very large specific-heat coefficient $\gamma_0$\,=\,5.65 J\,mol$^{-1}$ K$^{-2}$ characterize YbNi$_4$Mg as a typical SHF compound. Its SHF state intertwines with an antiferromagnetic short-range ordering below 0.3 K, where a large Wilson ratio $R_W$\,=\,32.1 is observed. Markedly, an enhanced Wilson ratio appears in a number of SHF materials lacking long-range magnetic order. 
Given that SHF compounds are generally characterized by weak Kondo coupling and bear significant similarities to insulating quantum magnets with uncompensated local moments, the enhanced Wilson ratios are considered a fingerprint of residual spin dynamics. Our MCE measurement on YbNi$_4$Mg shows a strong cooling effect surpassing GGG, revealing a temperature minimum in the quasi-adiabatic $T(B)$ trace near a metamagnetic-like crossover at $B^*$\,=\,0.1 T. In fact, such a metamagnetic crossover revealing significant magnetic cooling effect is generically expected for SHF materials, due to field-induced suppression of the SHF state and the entwined short-range magnetism. Our results further substantiate the proposal of using SHF for subKelvin cooling.

\section{Acknowledgments}
This work was supported by the National Natural Science Foundation of China (Grant Nos. 12141002, 52088101 and 12474147), the National Key R\&D Program of China (Grant Nos. 2022YFA1402200, 2021YFA0718700 and 2023YFA1406000). and the Chinese Academy of Sciences through the Project for Young Scientists in Basic Research (Grant No. YSBR-057), the Strategic Priority Research Program (Grant No. XDB33000000), and the Scientific Instrument Developing Project (Grant No. ZDKYYQ20210003). A portion of this work was carried out at the Synergetic Extreme Condition User Facility (SECUF) in Beijing.

\bibliographystyle{apsrev4-2}
\bibliography{Yb141_ref}

\end{document}